\documentclass[submit]{agujournal2019}
\usepackage{apacite}
\usepackage{url} 
\usepackage{lineno}
\usepackage{bm}
\usepackage{amsmath}
\usepackage{siunitx}
\usepackage{float}
\usepackage{mathtools}
\usepackage{cite}
\usepackage{natbib}
\usepackage{color}

\journalname{Space Weather}

\begin{document}

\title{A {Peculiar} ICME Event in August 2018 Observed with the Global Muon Detector Network}

\authors{W. Kihara\affil{1}, K. Munakata\affil{1}, C. Kato\affil{1}, R. Kataoka\affil{2}, A. Kadokura\affil{2}, S. Miyake\affil{3}, M. Kozai\affil{4}, T. Kuwabara\affil{5}, M. Tokumaru\affil{6}, R. R. S. Mendonça\affil{7}, E. Echer\affil{7}, A. Dal Lago\affil{7}, M. Rockenbach\affil{7}, N. J. Schuch\affil{7}, J. V. Bageston\affil{7}, C. R. Braga\affil{8}, H. K. Al Jassar\affil{9}, M. M. Sharma\affil{9}, M. L. Duldig\affil{10}, J. E. Humble\affil{10}, P. Evenson\affil{11}, I. Sabbah\affil{12}, and J. K\'ota\affil{13}}

\affiliation{1}{Physics Department, Shinshu University, Matsumoto, Japan}
\affiliation{2}{National Institute of Polar Research, Tachikawa, Japan}
\affiliation{3}{Department of Electrical and Electronic Systems Engineering, National Institute of Technology, Ibaraki College, Japan}
\affiliation{4}{Institute of Space and Astronautical Science, Japan Aerospace Exploration Agency, Sagamihara, Japan}
\affiliation{5}{Graduate School of Science, Chiba University, Chiba City, Japan}
\affiliation{6}{Institute for Space-Earth Environmental Research, Nagoya University, Nagoya, Japan}
\affiliation{7}{National Institute for Space Research, São José dos Campos, Brazil}
\affiliation{8}{George Mason University, Fairfax, VA, USA}
\affiliation{9}{Physics Department, Kuwait University, Kuwait City, Kuwait}
\affiliation{10}{School of Natural Sciences, University of Tasmania, Hobart, Australia}
\affiliation{11}{Bartol Research Institute, Department of Physics and Astronomy, University of Delaware, Newark, DE, USA}
\affiliation{12}{Department of Applied Sciences, College of Technological Studies, Public Authority for Applied Education and Training, Shuwaikh, Kuwait}
\affiliation{13}{Lunar and Planetary Laboratory, University of Arizona, Tucson, AZ, USA}

\correspondingauthor{Wataru Kihara}{19ss204g@shinshu-u.ac.jp}

\vspace{10mm}
\begin{center}(accepted for publication in the \it{Space Weather})\end{center}

\begin{keypoints}
\item We derived the spatial distribution of cosmic rays associated with a peculiar ICME event that caused a large magnetic storm in August 2018.
\item We found a cosmic-ray density increase possibly resulting from the MFR compression by the following faster solar wind.
\item The Global Muon Detector Network observed this density increase as a macroscopic modification of this geoeffective flux rope.
\end{keypoints}

\begin{abstract}
We demonstrate that global observations of high-energy cosmic rays contribute to understanding unique characteristics of a large-scale magnetic flux rope causing a magnetic storm in August 2018. Following a weak interplanetary shock on 25 August 2018, a magnetic flux rope caused an unexpectedly large geomagnetic storm. It is likely that this event became geoeffective because the flux rope was accompanied by {a} corotating interaction region and compressed by high-speed solar wind following the flux rope. In fact, a Forbush decrease was observed in cosmic-ray data inside the flux rope as expected, and a significant cosmic-ray density increase exceeding the unmodulated level before the shock was also observed near the trailing edge of the flux rope. The cosmic-ray density increase can be interpreted in terms of the adiabatic heating of cosmic rays near the trailing edge of the flux rope, as {the corotating interaction region prevents} free expansion of the flux rope and results in the compression {near} the trailing edge. {A northeast-directed} spatial gradient in the cosmic-ray density was also {derived} during the cosmic-ray density increase, suggesting that the center of the heating near the trailing edge is located {northeast} of Earth. This is one of the best examples demonstrating that the observation of high-energy cosmic rays {provides us with information that can only be derived from the cosmic ray measurements to} observationally constrain the three-dimensional macroscopic picture of the interaction between coronal mass ejections and the ambient solar wind, which is {essential for prediction of} large magnetic storms.
\end{abstract}

\section{Introduction}
\label{seq:1}
Solar eruptions such as coronal mass ejections (CMEs) cause environmental changes in various ways in near Earth space. It is known that major geomagnetic storms can be triggered by the arrival of an interplanetary counterpart of  {a} CME (ICME) at Earth along with a strong {southward} interplanetary magnetic field (IMF){,} which allows solar wind energy and plasma to enter the magnetosphere. {A} magnetic flux rope (MFR), which is often observed in an ICME with magnetic field lines winding about the central axis, is recognized as a key factor making an ICME such a powerful driver of an intense space weather storm. While ICMEs accompanied by a strong interplanetary shock (IP-shock) in a fast solar wind have attracted attention as geoeffective storms, the interaction of {moderate or slower ICMEs} with ambient solar wind structure and the interaction among a series of CMEs also play an essential role in producing an ICME causing a larger-than-expected magnetic storm\citep{Dal06,Liu14,Kata15}.\\

On its course in interplanetary space, an ICME driving {a} strong IP-shock forms a depleted region of the galactic cosmic rays (GCRs) behind the shock. When Earth enters this depleted region, cosmic-ray detectors at Earth's orbit detect a decrease of GCR intensity{,} which is known as a Forbush Decrease (FD) after {S. E. Forbush} \citep{Forb37}. The IP-shock accompanied by a turbulent magnetic sheath {inhibits} GCR transport into the inner heliosphere and sweeps GCRs away from Earth's orbit. The MFR behind the magnetic sheath, rapidly expanding in interplanetary space after the eruption from the Sun, also reduces GCR density inside the MFR by adiabatic cooling. At the same time, the GCR depletion either behind {the} IP-shock or in {the} MFR promotes the inward diffusion of GCRs. Due to the closed-field-line configuration of {the} MFR (in which both ends of each field line are anchored on the solar surface), GCRs {enter the MFR through} {drift and/or} cross-field diffusion{, the latter of} which is largely suppressed in {the} highly ordered strong IMF in {the} MFR even for high-energy particles.\\ 

By modeling the local part of {an} MFR with a straight cylinder, \citet{Muna06} numerically solved the GCR transport equation and found that the spatial distribution of GCR density in {MFRs} rapidly reaches a stationary state due to the balance between adiabatic cooling and inward cross-field diffusion. By assuming {an} axisymmetric straight cylinder for individual MFR{s}, \citet{Kuwa04} and \citet{Kuwa09} successfully derived from the observed GCR data the orientation and geometry of each MFR that were consistent with {in-situ} observations of IMF and the interplanetary scintillation (IPS) observations \citep{Toku07}. This demonstrates that cosmic-ray observations provide a useful tool for space weather studies \citep{Marl14}. In this paper, we study a particular ICME event observed in August, 2018 by analyzing the cosmic-ray data from the Global Muon Detector Network (GMDN).

\section{Overview of August, 2018 event}
\label{seq:2}
Figure 1 summarizes solar wind parameters measured in an ICME over four days between 24 and 27 August, 2018 (\url{https://omniweb.gsfc.nasa.gov/ow.html}). {Both the magnetic field and plasma data are observed by the} {\textit{Wind}} {spacecraft and time-shifted to Earth's location.} According to the list by Richardson and Cane ({column ``o'' in} \url{http://www.srl.caltech.edu/ACE/ASC/DATA/level3/icmetable2.htm}), this ICME event is caused by a CME eruption recorded at 21:24~UT on {20 August} by the LASCO coronagraphs on board the \textit{SOHO} satellite. {Following a weak IP-shock recorded at 03:00~UT on 25 August (see the pink vertical line in Figure 1),} {the sheath period can be identified by the enhanced fluctuation of IMF (a period delimited by the pink and the first} {blue} {vertical lines of about 12 hours after the IP-shock). After the sheath period,}  a significant enhancement of the IMF magnitude is recorded until {09:09}~UT on 26 August (panels a and c) in association with a systematic rotation of IMF orientation (panel d) indicating Earth's entrance into the MFR. {Following \citet{Chen19}, we define the MFR period as a period between 14:10 UT on 25 August and 09:09 UT on 26 August, delimited by a pair of} {blue} {vertical lines in Figure 1.}\\

A significant {southward} field is recorded in {the MFR} causing a gradual decrease of the {$D_{ST}$} index of geomagnetic field down to the minimum of -174 nT at 06:00 on {26 August} (panel e)(\url{http://wdc.kugi.kyoto-u.ac.jp/index.html}). Following the MFR period  {showing} the clear rotation of IMF orientation in Figure 1d, the gradual increase of solar wind speed is recorded along with significant fluctuations of IMF magnitude and orientation. We also note in Figure 1d that the IMF sector polarity is \textit{toward} in the period before the IP-shock as indicated by the GSE-longitude of IMF orientation (B$_{\rm{GSE}}$-long) around $300^{\circ}$, while it is \textit{away} after the MFR period as indicated by B$_{\rm{GSE}}$-long around $150^{\circ}$. This implies that this storm also {may involve}  heliospheric current sheet(s).\\

This event occurred in 2018 close to the solar activity minimum of solar cycle 24. {The CME was relatively slow, and occurred in slow solar wind, taking about five days to arrive at Earth after the CME eruption on the Sun. The} solar wind velocity enhancement after the IP-shock is also weak and seems to be insufficient to cause the large solar wind compression and significant enhancement of the {southward} IMF {that} triggered a major geomagnetic storm. \citet{Chen19} attributed peculiarities of this storm to the MFR compression by the following faster solar wind and \citet{Dal06} also presented a similar idea of MFR compression for an event {that} occurred in October 1999.\\

In this paper, we analyze the directional anisotropy of high-energy GCRs observed during this event. Since the GCR anisotropy arises from the diffusion and drift streamings{, which are} both proportional to the spatial gradient of GCR density, we can deduce from the observed anisotropy the three dimensional spatial distribution of GCRs which reflects the average magnetic field geometry extending over the large scale comparable to Larmor radii of high-energy GCRs in the IMF. {Our derivation of the GCR density gradient is based on the observational finding by \citet{Bieb98} that the drift is a primary source of  {the} ICME-related anisotropy observed with neutron monitors. We observed this with the higher rigidity response of GMDN and it has been recognized that the GCR density gradient derived from the observed anisotropy is rather insensitive to assumptions for the parallel and perpendicular diffusions. As already shown in a series of our papers, this allowed us to deduce from the observed anisotropy the orientation of cosmic ray density minimum viewed from Earth.} Readers can find examples of such analyses in \citet{Marl14} and references therein.\\

\begin{figure}[H]
\begin{center}
\includegraphics[natwidth=15cm,natheight=17cm]{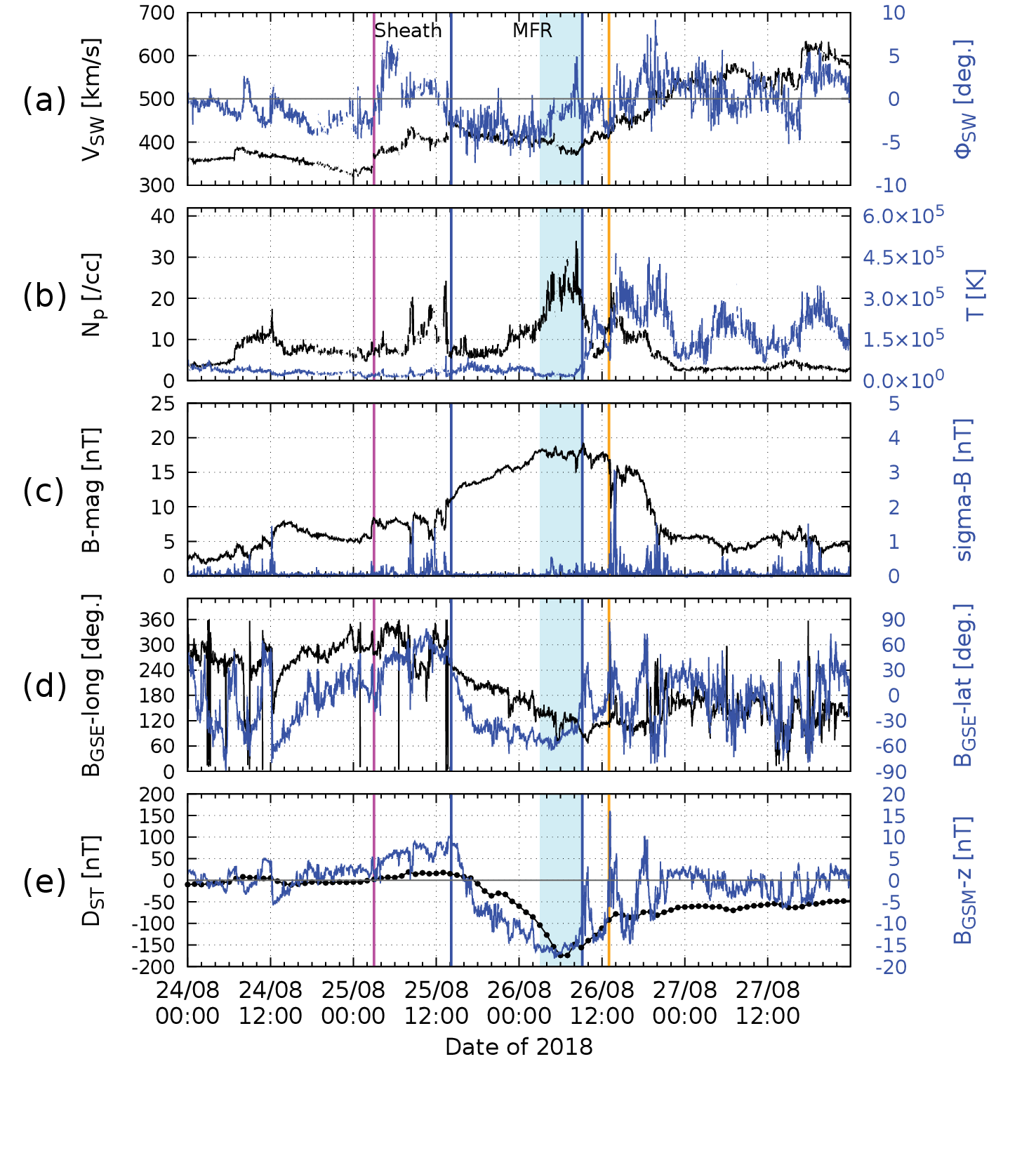}
\end{center}
\caption{Solar wind parameters and {$D_{ST}$} index {for 24-27 August}, 2018. From top to bottom, each panel shows one minute solar wind parameters; (a) magnitude of solar wind velocity (black curve) and ``flow angle'' of solar wind ($\phi_{SW}=\tan^{-1}(V_y/|V_x|$) (blue curve), (b) proton density (black) and temperature (blue), (c) IMF magnitude (black) and its fluctuation (blue), (d) GSE-longitude (black) and latitude (blue) of IMF orientation, (e) GSM-z component of IMF (blue) and hourly value of the {$D_{ST}$} index (black). The pink vertical line indicates the timing of IP-shock identified by the shock of IMF at {03:00}~UT on {25 August}, while a pair of {blue} vertical lines delimit the MFR period reported by \citet{Chen19}. The blue shaded area indicates six hours between 03:00~UT and 09:00~UT on {26 August} when an increase is observed in the cosmic-ray density (see Figure 2a and {Section} 4). The orange vertical line indicates the second stream interface at 13:00 UT on {26 August. As indicated at the top of the figure, we define the ``MFR period'' delimited by a pair of} {blue} {vertical lines and the ``sheath period'' between the pink and the first} {blue} {vertical lines (see text).}}
\label{fig:1}
\end{figure}

\begin{figure}[H]
\begin{center}
\includegraphics[natwidth=15cm,natheight=15cm]{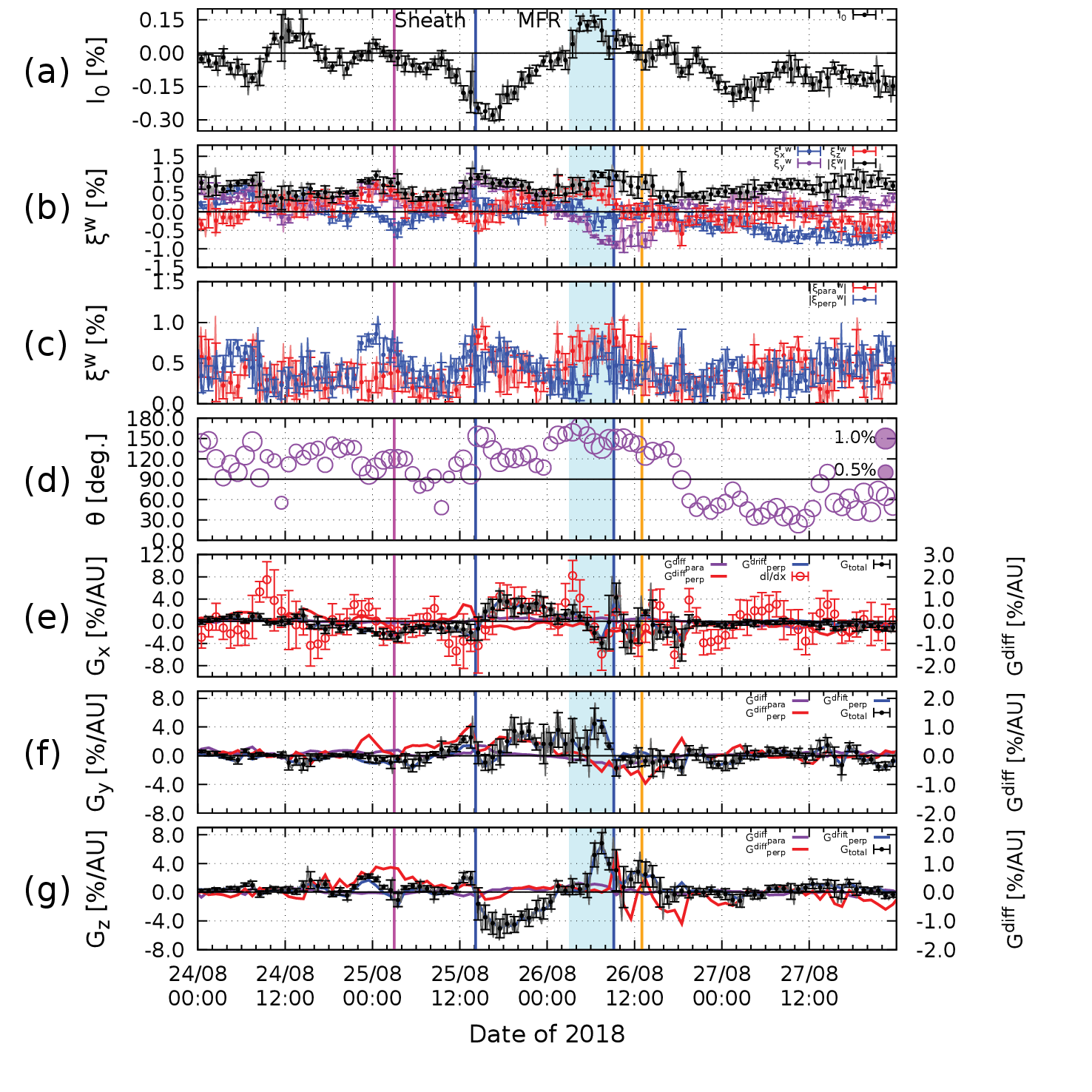}
\end{center}
\caption{Cosmic-ray density, anisotropy and density gradient at {60 GV} derived from GMDN data {for 24-27 August}, 2018. Each panel displays; (a) cosmic-ray density $I_0(t)$, (b) magnitude (black curve) and GSE-$x$ (blue), -$y$ ({purple}) and -$z$ (red) components of the anisotropy vector $\bm{\xi}^w(t)$ in the solar wind frame, (c) magnitudes of components of $\bm{\xi}^w(t)$ parallel (red) and perpendicular (blue) to IMF, {(d) a bubble plot showing the magnitude of hourly anisotropy vector $|\bm{\xi}^w(t)|$ as a function of the pitch angle ($\theta$) between $\bm{\xi}^w(t)$ and IMF vector,} while  (e)-(g) three GSE components of the density gradient vector $\bm{G}(t)$. {The area} of each circle in (d) is proportional to $|\bm{\xi}^w(t)|$ as indicated in the legend in the right top corner of the panel. {In each of panels (e)-(g), the contribution from the drift represented by the last term of Eq.\eqref{eq:G} is shown by the blue curve on the left vertical axis together with the total gradient component (black solid circles), while contributions from the parallel and perpendicular diffusions represented by the first and second terms of Eq.\eqref{eq:G} are shown by purple and red curves on the right vertical axis, respectively (note the scale of the right vertical axis is expanded four times the left axis).}

Each hourly value and error are deduced from the average and dispersion of 10-minute values in the corresponding one hour shown by a thin curve, respectively. {Open red circles in panel (e) show $\frac{1}{V_{SW}}\frac{dI_0(t)}{dt}$ calculated with $V_{SW}=400$ km/s for a test of $\bm{G}(t)$ derived from $\bm{\xi} ^w(t)$ (see text).} {The blue} shaded area indicates six-hours between 03:00~UT and 09:00~UT on {26 August} when the increase of $I_0(t)$ is observed in panel (a). The pink, {blue} and orange vertical lines are same as Figure 1. All data used for producing this figure are available in the Supporting Information (S2).}
\label{fig:2}
\end{figure}

\section{Cosmic-ray data and analyses}
\label{seq:3}
\subsection{Global Muon Detector Network (GMDN)}
The GMDN, which is designed for accurate observation of the GCR anisotropy, comprises four multidirectional muon detectors, ``Nagoya'' in Japan, ``Hobart'' in Australia, ``Kuwait City'' in Kuwait and ``S\~ao  Martinho da Serra'' in Brazil, recording muon count rates in 60 directional channels viewing almost the entire sky around Earth. Basic characteristics of directional channels of the GMDN are also available in the Supporting Information (S1). The median rigidity ($P_{m}$) of primary GCRs recorded by the GMDN, which we calculate by using the response function of the atmospheric muons to the primary GCRs given by numerical solutions of the hadronic cascade in the atmosphere \citep{Mura79}, ranges from about 50 GV for the vertical directional channel to about 100 GV for the most inclined directional channel, while the asymptotic viewing directions (corrected for geomagnetic bending of cosmic-ray orbits) at $P_{m}$ covers the asymptotic viewing latitude ($\lambda_{\rm asymp}$) from $72^{\circ}$N to $77^{\circ}$S. The representative $P_{m}$ of the entire GMDN is about 60 GV.

\subsection{Derivation of the GCR density and anisotropy}
We analyze the percent deviation of {the} 10-minute muon count rate $I_{i,j}(t)$ from {an average over 27 days between 12 August and 7 September}, 2018 in the $j$-th directional channel of the $i$-th detector ($i=1$ for Nagoya, $i=2$ for Hobart, $i=3$ for Kuwait and $i=4$ for São Martinho da Serra) in the GMDN at universal time $t$, after correcting for {local} atmospheric pressure and temperature effects. For our correction method of the atmospheric effects using the on-site measurement of pressure and the mass weighted temperature from the vertical profile of the atmospheric temperature provided by the Global Data Assimilation System (GDAS) of the National Center for Environmental Prediction, readers can refer to {\citet{Rafa16}}.\\

Since the observed temporal variation of $I_{i,j}(t)$ at the universal time $t$ includes contributions from variations of the GCR density (or ominidirectional intensity) $I_0(t)$ and anisotropy vector $\bm{\xi }(t)$, it is necessary to analyze each contribution separately. An accurate analysis of $I_0(t)$ and $\bm{\xi }(t)$ is possible only with global observations using multidirectional detectors. For such analyses, we model $I_{i,j}(t)$ in terms of $I_0(t)$ and three components ($\xi^{\rm GEO}_{x}(t), \xi^{\rm GEO}_{y}(t), \xi^{\rm GEO}_{z}(t)$) of $\bm{\xi }(t)$ in a geocentric (GEO) coordinate system, as
\begin{linenomath*}
\begin{eqnarray}
   I^{fit}_{i,j}(t) = I_0(t)c_{0 i,j}^0 &+& \xi^{\rm GEO}_x(t)(c_{1 i,j}^1 \cos \omega t_i - s_{1 i,j}^1 \sin \omega t_i) \nonumber \\
   &+& \xi^{\rm GEO}_y(t)(s_{1 i,j}^1 \cos \omega t_i + c_{1 i,j}^1 \sin \omega t_i) \nonumber \\
   &+& \xi^{\rm GEO}_z(t) c_{1 i,j}^0,
   \label{eq:fit}
\end{eqnarray}
\end{linenomath*}
where $t_{i}$ is the local time in hours at the $i$-th detector, $c^{0}_{0 i,j}$, $c^{1}_{1 i,j}$, $s^{1}_{1 i,j}$ and $c^{0}_{1 i,j}$ are coupling coefficients {which relate (or ``couple'') the observed intensity in each directional channel with the cosmic ray density and anisotropy in space} and $\omega=\pi/12$. In the GEO coordinate system, we set the $x$-axis to the anti-sunward direction in the equatorial plane, the $z$-axis to the geographical north perpendicular to the equatorial plane and the $y$-axis completing the right-handed coordinate system. The coupling coefficients in {Eq.}\eqref{eq:fit} are calculated by using the response function of the atmospheric muon intensity to primary GCRs \citep{Mura79} and given in the Supporting Information (S1). Note that the anisotropy vector $\bm{\xi }(t)$ in {Eq.}\eqref{eq:fit} is defined to direct opposite to the GCR streaming, pointing toward the upstream direction of the streaming (see also {Eq.}\eqref{eq:xi} in the next section). We derive the best-fit set of four parameters $\left( I_0(t), \xi^{\rm GEO}_{x}(t), \xi^{\rm GEO}_{y}(t), \xi^{\rm GEO}_{z}(t) \right)$ by solving the following linear equations.
\begin{linenomath*}
\begin{eqnarray}
\frac{\partial \chi^2}{\partial I_0(t)}=\frac{\partial \chi^2}{\partial \xi^{\rm GEO}_{x}(t)}=\frac{\partial \chi^2}{\partial \xi^{\rm GEO}_{y}(t)}=\frac{\partial \chi^2}{\partial \xi^{\rm GEO}_{z}(t)}=0,
   \label{eq:lsm}
\end{eqnarray}
where $\chi^2$ is the residual value of fitting defined, as
\begin{eqnarray}
\chi^2=\sum_{i,j}{(I_{i,j}(t)-I^{fit}_{i,j}(t))^{2}/\sigma_{ci,j}^2}
   \label{eq:res}
\end{eqnarray}
\end{linenomath*}
with $\sigma_{ci,j}$ denoting the count rate error of $I_{i,j}(t)$. The best-fit anisotropy vector {\boldmath{$\xi$}}$^{\rm GEO}(t)$ in the GEO coordinate system is then transformed to {{\boldmath{$\xi$}}$^{\rm GSE}(t)$} in the geocentric solar ecliptic (GSE) coordinate system for comparisons with the solar wind and IMF data.\\

{Eq.\eqref{eq:fit} does not include contributions from the second order anisotropy such as the bidirectional counter-streaming sometimes observed in the MFR in MeV electron/ion intensities. We also performed best-fit analyses adding five more best-fit parameters in Eq.\eqref{eq:fit} necessary to express the second order anisotropy and actually found an enhancement of the second order anisotropy in the MFR. However, we verified that the inclusion of the second order anisotropy} {does not change the obtained $I_0(t)$ and $\bm{\xi }(t)$ significantly keeping conclusions of the present paper unchanged.} {In this paper, therefore, we analyze only $I_0(t)$ and $\bm{\xi }(t)$ derived from Eq.\eqref{eq:fit}. We will present our analyses and discussion of the second-order anisotropy elsewhere.}\\

\subsection{Derivation of the spatial gradient of GCR density}
{Diffusive} propagation of GCRs in the heliosphere is described by the following transport equation {\citep{Park65, Glee69}}. 
\begin{linenomath*}
\begin{equation}
\frac{\partial U}{\partial t}+\bm{\nabla}\cdot\bm{S}=-\frac{\partial}{\partial p}(\frac{1}{3}p\bm{V}_{SW}\cdot\bm{\nabla}U),
\label{eq:cd}
\end{equation}
\end{linenomath*}
where $U(\bm{r},p,t)$ is the GCR density at position $\bm{r}$, momentum $p$ and time $t$, $\bm{V}_{SW}$ is the solar wind velocity. $\bm{S}(\bm{r},p,t)$ in {Eq.}\eqref{eq:cd} is the GCR streaming vector consisting of the solar wind convection and the diffusion terms, as
\begin{linenomath*}
\begin{equation}
\bm{S}=CU\bm{V}_{SW}-\bm{\kappa}\cdot\bm{\nabla}U
\label{eq:s}
\end{equation}
\end{linenomath*}
where $\bm{\kappa}$ is the diffusion tensor and $C$ is the Compton-Getting (CG) factor denoted by $C={1-\frac{1}{U}\frac{\partial}{\partial p}(\frac{1}{3}pU)}=(2+\gamma )/3$ {with an assumption of $U$ proportional to $p^{-\gamma}$ with the power-law index} $\gamma=2.7$. {The diffusion and drift anisotropy $\bm{\xi^D}$ is given as}
\begin{linenomath*}
\begin{equation}
{\bm{\xi^D}(t)}\equiv-\frac{3 \bm{S}}{v U}=\frac{3}{v}(\bm{\kappa}\cdot\bm{G}-C\bm{V}_{SW})
\label{eq:xi}
\end{equation}
\end{linenomath*}
where $v$ is the speed of GCR particle, which is approximately equal to the speed of light $c$, and $\bm{G}=\bm{\nabla}U/U$ is the spatial gradient of GCR density.\\

{We transform the observed anisotropy ${\bm{\xi}}^{\rm GSE}(t)$} by subtracting the solar wind convection and an apparent anisotropy arising from Earth's orbital motion around the Sun, as
\begin{linenomath*}
\begin{align}
\label{eq:xiw}
\bm{\xi}^w(t) =~ {{\bm{\xi}}^{\rm GSE}(t)}+(2+\gamma )\frac{\bm{V}_{SW}(t)-\bm{v}_E}{c}
\end{align}
\end{linenomath*}
where $\bm{v}_E$ is the velocity of Earth (30 km/s toward the orientation opposite to the GSE-y orientation). {We replace $\bm{\xi}^w(t)$ with $\bm{\xi}^D(t)$ as
\begin{linenomath*}
\begin{align}
\label{eq:xi2}
\bm{\xi}^w(t) =~\bm{\xi}^D(t)
\end{align}
\end{linenomath*}
by ignoring contribution to $\bm{\xi}^w(t)$ from other possible non-diffusion/drift anisotropy such as recently reported by \citet{Tort18} from the observation in a MFR\citep{Krit09}. Then, we can deduce the density gradient $\bm{G}$ by solving Eq.\eqref{eq:xiw} and Eq.\eqref{eq:xi2} for $\bm{G}$ as}
\begin{linenomath*}
\begin{align}
\label{eq:G}
\bm{G}(t) &= \frac{1}{R_L(t) \alpha _{\parallel}} \bm{\xi}^w_{\parallel}(t)+{\frac{\alpha_{\perp}}{R_L(t)(1+\alpha ^2_{\perp})}\bm{\xi}^w_{\perp}(t) +\frac{1}{R_L(t)(1+\alpha ^2_{\perp})} \frac{\bm{B}(t)}{B(t)} \times \bm{\xi}^w_{\perp}(t)}
\end{align}
\end{linenomath*}
where $R_L(t)=\frac{P}{c|\bm{B}(t)|}$ is the Larmor radius of particles with rigidity $P$ in magnetic field $\bm{B}(t)$ and $\bm{\xi}^w_{\parallel}(t)$ and $\bm{\xi}^w_{\perp}(t)$ are components of $\bm{\xi } ^w(t)$ parallel and perpendicular to $\bm{B}(t)$, respectively \citep{Koza16}. $\alpha_{\parallel}$ and $\alpha_{\perp}$ in {Eq.}\eqref{eq:G} are mean-free-paths of parallel and perpendicular diffusions, respectively, normalized by $R_L(t)$, as
\begin{linenomath*}
\begin{align}
\label{eq:alp1}
\alpha_{\parallel} = {\lambda _{\parallel}(t) /R_L(t)}\\
\label{eq:alp2}
\alpha_{\perp} = {\lambda_{\perp}(t)/R_L(t)}.
\end{align}
\end{linenomath*}
{According to current understanding that GCRs at neutron monitor and muon detector energies are in the ``weak-scattering'' regime \citep{Bieb04}, we assume $\lambda_{\perp}(t) \ll \lambda _{\parallel}(t)$. {Following} models widely used in the study of the large-scale GCR transport in the heliosphere \citep{Wibb98, Miya17},} we assume constant $\alpha_{\perp}=0.36$ for a period outside {the} MFR in this paper. {We also assume $\lambda _{\parallel}=1.9$ AU for the entire period}. For 60 GV cosmic rays in $|\bm{B}(t)|\sim5~\rm{nT}$ average magnetic field, $R_L$ is 0.27~AU resulting in $\lambda _{\perp}=0.096~\rm{AU}$ {and $\alpha_{\parallel}=7.2$}. For a period inside the MFR where the magnetic field is exceptionally strong, we use a constant $\lambda _{\perp}=0.010~\rm{AU}$ without changing {$\lambda _{\parallel}$}. Note that this $\lambda _{\perp}$ was obtained as an upper limit by \citet{Muna06}.\\

{We are aware that our ad-hoc assumptions of $\lambda _{\parallel}(t)$ and $\lambda _{\perp}(t)$ above are difficult to validate directly from observations. However, it will be shown in the next section that $\bm{G}(t)$ derived from the observed anisotropy $\bm{\xi}^w(t)$ in Eq.\eqref{eq:G} is significantly dominated by the contribution from the drift anisotropy represented by the last term on the right-hand side of Eq.\eqref{eq:G} and is insensitive to our ad-hoc assumptions of $\lambda _{\parallel}(t)$ and $\lambda _{\perp}(t)$.}\\

\section{Results}
\label{seq:4}
Figure 2 shows the GCR density ($I_0(t)$ in panel a), anisotropy ($\bm{\xi } ^w(t)$ in panels b-{d}) and density gradient ({$\bm{G}(t)$} in e-g ) {at 60 GV} obtained from our analyses of the GMDN data described in the preceding section using the solar wind velocity $\bm{V}_{SW}(t)$ and IMF $\bm{B}(t)$ in Figure 1. While we derived the best-fit parameters in {Eq.}\eqref{eq:fit} in every 10-minute interval, in this paper we only use the hourly average of six 10-minute values, because one hour is much shorter than the time scale ($R_{L}/V_{SW}\sim$~9~hours) for the solar wind to travel across the Larmor radius ($R_{L}$=0.089~AU) of 60~GV GCRs in IMF ($B\sim$~15~nT) with the average velocity ($V_{SW}\sim$~400~km/s) and enough for analyzing the spatial distribution of 60~GV GCRs. We also {calculated} the error of {the} hourly value of each parameter from the dispersion of 10-minute values. All data used for producing this figure are given in the Supporting Information (S2).\\

{Besides the random error of the best-fit parameters in Figure 2, there are possible sources of systematic error. For instance, the atmospheric effect results in the day-to-day offset of $I_{i,j}(t)$ which is almost the same for all directional channel in one detector, but generally different between detectors at different locations. We corrected for the effect by using the barometric and temperature coefficients ($\beta$ and $\alpha$ in the Supporting Information (S1)) derived in August 2018, instead of using nominal (or average) coefficients derived from the long-term observations. We verified that the local effect in $I_{i,j}(t)$ is significantly reduced with smaller $\chi^2$ in this way.} {Another source of systematic error is the second order anisotropy which is not included in Eq.\eqref{eq:fit}, but we verified that the inclusion of the second order anisotropy does not change the obtained $I_0(t)$ and $\bm{\xi }(t)$ significantly, as mentioned in the preceding section. We conclude, therefore, that systematic error is similar to or smaller than the random error in Figure 2.}\\

The cosmic-ray density $I_0(t)$ in Figure 2a starts decreasing {a few hours before} the IP-shock early on {25 August} and goes to the minimum of -0.28 \% at 16:30 UT and recovers to the level before the IP-shock early on {26 August}. This is a well-known feature of a moderate Forbush decrease indicating Earth's entrance into the cosmic-ray depleted region formed behind the shock and in the magnetic flux rope (MFR). {A small local maximum is seen in $I_0(t)$ at around 10:00 UT of 25 August being superposed on the gradual decrease in the sheath period. We think that this is probably due to the discontinuity recorded in the IMF magnitude and longitude seen in Figures 1c-d, because such a local increase of $I_0(t)$ is often observed by the GMDN at the IMF discontinuity \citep{Muna18}.}\\

A marked feature of this event is a ``hump'' {in} the density in which $I_0(t)$ increased to the maximum at around 06:00 UT on {26 August} exceeding the unmodulated level before the shock. As discussed later, this increase {is likely} caused by the compression {of the trailing edge of the MFR by the faster solar wind following the MFR} (see the blue shaded period in Figure{s} 1 and 2). Figure 2a also shows a significant variation of $I_0(t)$ around noon on {24 August} when the disturbance is recorded in the solar wind parameters. We {do not} analyze this variation in this paper, but \citet{Abun20} considered it as an indication of another {I}CME which they identified.\\

Figure 2c shows magnitudes of $\bm{\xi}^w_{\parallel}(t)$ and $\bm{\xi}^w_{\perp}(t)$, the parallel and perpendicular components of the anisotropy in {Eq.}\eqref{eq:G}, while Figure 2d displays $|\bm{\xi } ^w(t)|$ as a function of the pitch angle {($\theta$)} between $\bm{\xi } ^w(t)$ and the IMF. {During a period between $\sim$12:00 UT on 25 August and $\sim$0:00 UT on 26 August in the MFR period when $I_0(t)$} {recovers from its minimum} {and the amplitude of $\bm{\xi}^w(t)$ increases,} the perpendicular component ($\bm{\xi}^w_{\perp}(t)$) exceeds the parallel component ($\bm{\xi}^w_{\parallel}(t)$) except for a few hours prior to the minimum of $I_0(t)$. {This indicates the dominant contributions from drift anisotropy inside the magnetic flux rope.}\\

{During a few hours after $\sim$03:00~UT on 26 August in the blue shaded period, on the other hand, $\bm{\xi}^w(t)$ is dominated by the parallel anisotropy. {\citet{Tort18} reported a strong anisotropy parallel to the magnetic field observed in a MFR, as predicted from a theory by \citet{Krit09}. Such parallel anisotropy cannot be expressed by $\bm{\xi}^D(t)$ in Eq.\eqref{eq:xi} based on the diffusion and drift picture particularly in a MFR where $\lambda _{\parallel}(t)$ can be comparable to or even longer than the scale size of the MFR. The theory predicts a parallel anisotropy inside MFRs, due to an inflow of cosmic rays along one leg of the MFR, caused by guiding center drifts, and an outflow along the other. Cosmic rays enter the MFR along a leg of the MFR where the field line winding is counterclockwise (viewing from the wide part of the field line cone) and outward along the other leg with clockwise winding \citep{Krit09}. Since \citet{Chen19} reported that the MFR in Figure 1 is left-handed and its mean magnetic field directs southward, the theory predicts cosmic rays to enter the MFR at the southern leg and exit at the northern leg, resulting in a negative $\xi^w_z$ (north directing flow) at Earth's orbit. However, the observed $\xi^w_z$ in Figure 2b (red curve) is positive (south directing flow) during the blue shaded period, in contradiction to the prediction. We conclude therefore that the contribution from such unidirectional parallel flow to the observed anisotropy is not dominant in this particular event, even if it exists.}\\

Figures 2e-{g} show three GSE-components of the density gradient ($\bm{G}(t)$) {displayed by black curves. {Blue curves in panels (e)-(g) show contributions to $\bm{G}(t)$ from the drift represented by the last term of Eq.\eqref{eq:G} on the left vertical axis, while purple and red curves display contributions from the parallel and perpendicular diffusions  represented by the first and second terms, respectively, on the right vertical axis.} It is clear that the derived $\bm{G}(t)$ is significantly dominated by the contribution from the drift anisotropy and contributions from parallel and perpendicular diffusions are small (Bieber and Evenson, 1998). {By changing assumptions of $\lambda _{\parallel}(t)$ and $\lambda _{\perp}(t)$ in Eq.\eqref{eq:G} each in a wide range}, we also verified that the derived $\bm{G}(t)$ is insensitive to these parameters. {For instance, we calculated $\bm{G}(t)$ by assuming a constant $\lambda _{\perp}(t)=0.010$ AU during the MFR period between the two vertical blue lines and verified that the difference of $\bm{G}(t)$ from the black solid circles in panels (e)-(g) are well within errors.} As already shown in a series of our papers, this allowed us to deduce from the observed anisotropy the CME geometries viewed from Earth successfully in accordance with other observations (\citealp{Kuwa04}, \citeyear{Kuwa09}).}\\

{As a test of our $\bm{G}(t)$ derived from $\bm{\xi } ^w(t)$, we also calculate $G_{x}$ from $\frac{1}{V_{SW}}\frac{dI_0(t)}{dt}$ which is expected to be observed at Earth in the case of the stationary $G_{x}$ passing Earth with the solar wind velocity $V_{SW}$. Red open circles superposed in Figure 2e show $\frac{1}{V_{SW}}\frac{dI_0(t)}{dt}$ calculated with $V_{SW}=400$ km/s. It is seen that $G_x(t)$ derived from $\bm{\xi } ^w(t)$ shown by black solid circles is consistent within errors with $\frac{1}{V_{SW}}\frac{dI_0(t)}{dt}$ independently derived from $I_0(t)$, particularly during the MFR period including the blue shaded period. This supports the validity of our $\bm{G}(t)$ derived from $\bm{\xi}^w(t)$.}\\

{It is seen in the black curves in Figures 2e-f that $G_{x}(t)$ and $G_{y}(t)$ change their signs from negative to positive around the time of the minimum $I_0(t)$ in the MFR period, while $G_{z}(t)$ remains negative. This is qualitatively consistent with the MFR orientation (the elevation angle of the MFR is about $-51^\circ$ and the azimuthal angle is about $299^\circ$) and the Grad-Shafranov plot presented in \citet{Chen19}, indicating that the cosmic-ray density minimum region formed along the MFR axis passed north of Earth approaching from the sunward direction  and then leaving \citep{Kuwa04}.}\\

Another notable {feature}{s} of $\bm{G}(t)$ are positive enhancements of $G_{z}(t)$ and $G_{y}(t)$ {in the later MFR period} between 03:00 UT and 09:00 UT on {26 August} (blue shaded period in Figures 1 and 2) when the orientation of the strongest IMF turned {northeast after the maximum of $I_0(t)$ observed in the hump in Figure 2a}. This indicates that a region with higher $I_0(t)$ exists {northeast} of Earth ({$y_{GSE}>0$, $z_{GSE}>0$}) and a significant diffusion anisotropy from there is observed along the IMF.\\

Finally, we note a large variation of $G_{x}(t)$ around the maximum of $I_0(t)$ in the hump. In particular, $G_{x}(t)$ with a magnitude of 4\%/AU changes its sign from positive to negative in four hours. In the next section, we will discuss the physical origin of the hump of $I_0(t)$ and associated $\bm{G}(t)$.\\

\section{Discussions}
\label{seq:5}
As discussed in {Section} 2, the MFR in this event is followed by a gradual increase of the solar wind speed ($V_{SW}$) starting at $\sim$09:00 UT on {26 August}. This increase of $V_{SW}$ is consistent with a typical stream interface, as identified by the proton temperature increase, proton density decrease and{,} a negative-to-positive flip of flow angle seen in Figure 1 \citep{Kata06}. There are two major interesting points in this event. Firstly, stream interfaces are usually formed in corotating interaction regions, which clearly separate the slow and fast solar wind streams. In this event, however, the slow solar wind was replaced by a slow MFR. The existence of such a clear discontinuity at the trailing edge of {the} MFR therefore suggests a large-scale compression of the trailing part of {the} MFR by the following ambient solar wind. Secondly, a secondary enhancement is seen in the solar wind speed at $\sim$13:00~UT on {26 August as indicated by the orange vertical lines in Figures 1 and 2}, which again shows a stream interface-like variation, as identified by an increase of proton temperature and a decrease of proton density. Such doublet structures in the solar wind speed enhancement are often observed in corotating interaction regions in association with longitudinally elongated and complex coronal hole(s), as was also discussed in \citet{Abun20}. We also note in Figure 2a that the cosmic-ray density $I_0(t)$ starts decreasing after the MFR period. This is again consistent with cosmic-ray intensity variation observed in the corotating interaction regions where $I_0(t)$ peaks near the stream interface and then starts falling in the leading edge of the high-speed stream \citep{Rich04}.\\

In the preceding section, we reported a significant increase (a hump) of cosmic-ray density $I_0(t)$ observed in the GMDN data near the end of the MFR period. Recently, \citet{Abun20} have also analyzed cosmic-ray data observed by neutron monitors in August 2018. Neutron monitors have their maximum response to cosmic rays with $P_{m}\sim10$ GV which is $\sim1/5$ of $P_{m}$ for muon detectors. While they found no clear increase in $I_0(t)$, they reported large increases, called ``bursts'', in count rates of some neutron monitors near the trailing edge of the MFR and attributed these ``bursts'' to the enhancement of {$\xi_z(t)$} and the geomagnetic storm occurring at the same time \citep{Abun20}. {They did not present the density gradient ($\bm{G}(t)$) deduced from the observed anisotropy.} When the geomagnetic field is weakened during the storm and the geomagnetic cutoff rigidity ($P_c$) of cosmic rays is reduced, allowing more low-energy particles to reach the ground level detectors, the asymptotic viewing direction of a neutron monitor is changed due to the reduced magnetic deflection of cosmic-ray orbits in the magnetosphere. A similar idea was also presented by \citet{Moha16}  to interpret the ``cosmic-ray short burst'' observed by a muon detector in June 2015. Based on calculations of cosmic-ray trajectory in the latest model of geomagnetic field, however, analyses of the same event by \citet{Muna18} showed that the reductions of $P_c$ and the magnetic deflection of cosmic-ray orbits are not enough to cause the observed intensity increase of 60 GeV cosmic rays monitored by the G{M}DN, because the G{M}DN {only has a} small response to cosmic rays with rigidities around $P_c$. They attributed the burst to enhancements of $I_0(t)$ and $\bm{\xi}(t)$ outside the magnetosphere caused by Earth's crossing the heliospheric current sheet \citep{Muna18}.\\

The observation that, in the blue shaded region,  $I_0$ rose above its undisturbed level suggests that GCRs gained energy relative to the quiet period. A plausible cause of the energy gain may be a compression of the plasma occurring in the MFR. In Appendix A we outline our preliminary model considering this option. Our model considers the energy gain in a slab parallel to {the} $yz$ plane {including the magnetic field} with thickness $2d$ and perpendicular to the solar wind in $-x$-direction. The model also assumes that the global expansion of the MFR is not affected by the local compression because the expansion continues on the leading side of the MFR. This model quantitatively reproduces the observed time profiles of $I_0(t)$ and $G_{x}(t)$ during the blue shaded period by using the plasma density enhancement in Figure 1b as a proxy of the local compression rate (see Figure A1 in Appendix). However, it does not take into account transport to/from the remaining part of the MFR which is magnetically connected to the trailing edge. The model in Appendix A, therefore, would be unphysical if the MFR of interest is an ``ideal'' axisymmetric expanding cylinder such as those analyzed by \citet{Kuwa04} \citep{Muna06}. Figure 2, however, suggests that this event is quite peculiar and far from an ``ideal'' MFR.\\

By analyzing GMDN observations of 11 CME events using the expanding axisymmetric cylinder model, \citet{Kuwa09} showed that $I_0(t)$ typically starts decreasing after the beginning of the MFR period and reaches its minimum at about one half or one third of the MFR period when Earth passes the closest point to the MFR axis, while $G_x(t)$ changes its sign from negative to positive at the time. $I_0(t)$ in Figure 2a, however, reaches its minimum much earlier, only a few hours after the start of the MFR period and $G_x(t)$ in Figure 2e changes sign at the same time. $I_0(t)$ then starts recovering almost monotonically toward the peak in the blue shaded period. We think that these are indications of cosmic-ray heating (and cooling) in operation over an entire MFR which cannot be reproduced by the simple slab model in this paper.  The stark peculiarity of this MFR is also seen in Figure 1 and in the Grad-Shafranov plot in Figure 9 of \citet{Chen19} in which the MFR core is shifted close to the trailing edge being surrounded by tightly wound magnetic field lines. To fully understand the observed $I_0(t)$ and $\bm{G}(t)$ reflecting the modification of the MFR, therefore, a more practical and detailed model taking account of adiabatic heating/cooling together with parallel and perpendicular diffusions of cosmic rays in a three dimensional entire MFR is awaited. This is planned for our future investigation.\\
 
Dynamic evolution of the MFR, i.e. the simplest adiabatic heating of cosmic rays, is a possible cause of the cosmic-ray increase observed near the trailing edge of MFR in this paper, as discussed above{, although we do not know any literatures reporting the compressive heating of plasma or energetic ions/electrons inside an MFR near the trailing edge. On the other hand, there are other possibilities} to observe the cosmic-ray increase, because steady structures such as the heliospheric current sheet and corotating interaction region affect the drift of cosmic rays and can also modulate the large-scale {spatial distribution of cosmic-ray density} and anisotropy \citep{Oka08,Fusi10}. This study therefore provides an important clue to examine cosmic-ray diffusions parallel and perpendicular to the IMF, and the dependence on the IMF magnitude in great detail, via close collaborations with the drift-model simulations of cosmic-ray transport \citep{Miya17} and the cutting-edge MHD simulations \citep{Shio16,Matu19}. After all, the weak solar wind condition provided a unique opportunity to study the cosmic-ray increase in this paper.\\

{We learned from this study that the GMDN observed the cosmic ray density increase and the associated gradient as evidence of the MFR compression by the faster following solar wind which made this peculiar event geoeffective. This evidence is unique and independent of other observations including in-situ measurements and demonstrates the value of cosmic ray measurements in understanding the physics of space weather forecasting.}\\

\appendix
\section{Slab model of the adiabatic heating of cosmic rays}
{
In this section, we consider the cosmic ray transport in the blue shaded period in Figures 1 and 2 when a ``hump'' of $I_{0}(t)$ is observed in Figure 2a. We model the solar wind velocity} $\bm{V}_{SW}$ as
\begin{linenomath*}
\begin{equation}
\bm{V}_{SW}=\bar{\bm{V}}_{SW}+\bm{u}
\label{eq:A1}
\end{equation}
\end{linenomath*}
where $\bar{\bm{V}}_{SW}$ is the average velocity and $\bm{u}$ is an additional compression or expansion velocity. The observed temporal variation of $\bm{V}_{SW}$ during the blue shaded period in Figure 1a shows first a gradual decrease and then an increase near the end of period. This is qualitatively consistent with the compression velocity $\bm{u}$ directing inward of the blue shaded area on both sides of a point where $\bm{u}$ becomes zero, because such $\bm{u}$ increases (decreases) $\bm{V}_{SW}$ on the side of that point closer to (farther from) the Sun. The point where $\bm{u}$ becomes zero looks shifted to the later period probably due to $\bar{\bm{V}}_{SW}$ increasing (accelerating). By using the phase-space density of cosmic rays $f(\bm{r}, p, t)=U(\bm{r}, p, t)/(4\pi p^2)$, Eq.\eqref{eq:cd} is written as
\begin{linenomath*}
\begin{equation}
\frac{\partial f}{\partial t}=\bm{\nabla}\cdot(\bm{\kappa}\cdot\bm{\nabla}f)-\bm{u}\cdot\bm{\nabla}f+\frac{1}{3}(\bm{\nabla}\cdot\bm{u})p\frac{\partial f}{\partial p}.
\label{eq:A2}
\end{equation}
\end{linenomath*}
For $f(x, p, t)$ in one-dimension perpendicular to the mean magnetic field, this equation becomes
\begin{linenomath*}
\begin{equation}
\frac{\partial f}{\partial t}=\kappa_{\perp}\frac{\partial^2 f}{\partial x^2}-u\frac{\partial f}{\partial x}+\frac{1}{3}\frac{\partial u}{\partial x}p\frac{\partial f}{\partial p},
\label{eq:A3}
\end{equation}
\end{linenomath*}
where $x$ is the position measured from the center of the slab toward the Sun and $\kappa_{\perp}=\frac{1}{3}\lambda_{\perp}v$ is the spatially uniform perpendicular diffusion coefficient.
}\\

{
By assuming the self-similar compression of the slab, we set
\begin{linenomath*}
\begin{equation}
u(x,t)=-u_c\frac{x}{d(t)},
\label{eq:A4}
\end{equation}
\end{linenomath*}
where $d(t)$ is the half-thickness of the slab at time $t$ defined as
\begin{linenomath*}
\begin{equation}
d(t)=d_0(1-\frac{t}{t_r})
\label{eq:A5}
\end{equation}
\end{linenomath*}
with $d_0$ denoting $d(t)$ at $t=0$. $t_r (>0)$ is the reference time when $d(t)$ goes to zero, but we assume that this does not happen during the event period we discuss. The real importance of $t_r$ is that it is the inverse rate of relative compression giving the magnitude of constant compression velocity at $x=d(t)$ by $u_c=\vert\frac{dd(t)}{dt}\vert=\frac{d_0}{t_r}$. In this paper, we do not include the adiabatic cooling due to the large scale expansion of solar wind, which operates in longer time scale, but focus on the effect of local compression at $0<t\ll t_r$. By introducing (A4), we get (A3) as
\begin{linenomath*}
\begin{equation}
\frac{\partial f}{\partial t}=\kappa_{\perp}\frac{\partial^2 f}{\partial x^2}+u_c\frac{x}{d}\frac{\partial f}{\partial x}-\frac{u_c}{d}\frac{1}{3}p\frac{\partial f}{\partial p}.
\label{eq:A6}
\end{equation}
\end{linenomath*}
}\\

{
We replace $x$ with a dimensionless variable $s$, defined as
\begin{linenomath*}
\begin{equation}
s(x,t)=\frac{x}{d(t)},
\label{eq:A7}
\end{equation}
\end{linenomath*}
where $-1\leq s \leq1~ (-d\leq x \leq d)$ and we get an equation for $f_s(s, p, t)=f(x,p,t)$ as
\begin{linenomath*}
\begin{equation}
\frac{\partial f_s}{\partial t}=\frac{\kappa_{\perp}}{d^2}\frac{\partial^2 f_s}{\partial s^2}-\frac{u_c}{d}\frac{1}{3}p\frac{\partial f_s}{\partial p},
\label{eq:A8}
\end{equation}
\end{linenomath*}
by using
\begin{linenomath*}
\begin{equation}
\frac{\partial f}{\partial t}=\frac{\partial f_s}{\partial t}+\frac{\partial s}{\partial t}\frac{\partial f_s}{\partial s}=\frac{\partial f_s}{\partial t}+\frac{s}{d}u_c\frac{\partial f_s}{\partial s}.
\label{eq:A9}
\end{equation}
\end{linenomath*}
For $f_s$ in the steady state, we get
\begin{linenomath*}
\begin{equation}
\frac{\partial^2 f_s}{\partial s^2}=\frac{u_cd}{\kappa_{\perp}}\frac{1}{3}p\frac{\partial f_s}{\partial p}.
\label{eq:A10}
\end{equation}
\end{linenomath*}
}\\

{
We finally assume a single power-law dependence on $p$ for $f_s$ outside the slab, as
\begin{linenomath*}
\begin{equation}
f_s(s,p)=p^{-(2+\gamma)}F(s)
\label{eq:A11}
\end{equation}
\end{linenomath*}
with $\gamma$ denoting the exponent of the momentum spectrum ($U\propto p^{-\gamma}$) set equal to 2.7 and obtain the equation to be solved, as 
\begin{linenomath*}
\begin{equation}
\frac{d^2 F}{ds^2}=-\frac{2+\gamma}{3\kappa_0}F
\label{eq:A12}
\end{equation}
\end{linenomath*}
where $\kappa_0$ is a dimensionless parameter defined as
\begin{linenomath*}
\begin{equation}
\kappa_0=\frac{\kappa_{\perp}}{u_cd}.
\label{eq:A13}
\end{equation}
\end{linenomath*}
}\\

{
By assuming a uniform compression of the plasma with the density $N_p$ in the slab, which keeps $d(t)N_p$ constant, we replace $u_c$ in Eq. (A13) with $d(t)\frac{1}{N_p}\frac{dN_p}{dt}$ and get
\begin{linenomath*}
\begin{equation}
\kappa_0=\frac{\kappa_{\perp}N_p}{d^2\frac{dN_p}{dt}}=\frac{1}{3}\lambda_{\perp}v\frac{1}{d(t)^2\frac{d\log N_p}{dt}}
\label{eq:A14}
\end{equation}
\end{linenomath*}
where $\lambda_{\perp}$ is the meanfreepath of perpendicular diffusion and $v$ is the velocity of cosmic ray particles.} Eq. (A12) can be written using the spatial density gradient $G_x=\frac{1}{F}\frac{dF}{dx}$ with $x$ as
\begin{linenomath*}
\begin{equation}
-\frac{{(}2+\gamma{)}}{\lambda_{\perp}v}\frac{d\log N_p}{dt}=\frac{1}{F}\frac{d^2 F}{dx^2}=\frac{1}{F}\frac{d (FG_x)}{dx}=G_x^2+\frac{dG_x}{dx}\approx\frac{dG_x}{dx},
\label{eq:A15}
\end{equation}
\end{linenomath*}
because $G_x$ is less than 5\%/AU=$5\times10^{-2}$/AU at most and $G_x^2$ is much smaller than $\frac{\partial G_x}{\partial x}$ which is $\sim1$/AU$^2$ (see Figure A1b below). By integrating Eq. (A15) by $x$, we get
\begin{linenomath*}
\begin{equation}
G_x\approx-\frac{{(}2+\gamma{)}}{\lambda_{\perp}v}\frac{d\log N_p}{dt}x,
\label{eq:A16}
\end{equation}
\end{linenomath*}
and
\begin{linenomath*}
\begin{equation}
{I_0}\approx\frac{{(}2+\gamma{)}}{2\lambda_{\perp}v}\frac{d\log N_p}{dt}{(d^2-x^2)},
\label{eq:A17}
\end{equation}
where we set $I_0=0$ at $x=\pm d$ as a boundary condition.
\end{linenomath*}\\

The steady state solution of {$G_x$} in {Eq.}\eqref{eq:A16} is a linear function of the distance {$x=V_{SW}dt$ where $dt$} {is the time relative to the time of passage of the center of the slab.} In discussions here, we assume {the thickness of slab} is constant, ignoring its temporal variation during the blue shade period. {This is seen in Figure A1b showing the observed $G_x$ which can be fitted by a linear function of $x$ calculated with a constant $V_{SW}=400$ km/s and $dt$} {relative to} {05:16 UT on 26 August when the best-fit line crosses the horizontal axis.} The slope of {this} best-fit line is {$-1.47\times10^2 (\%/\rm{AU}^2)$.} By using $\lambda_{\perp}=$0.010 AU and $\gamma=2.7$ {in Eq.\eqref{eq:A16}}, we get $\frac{d\log N_p}{dt}=2.2\times10^{-2}$/hour, necessary for the heating, while the average $\frac{d\log N_p}{dt}$ calculated from hourly mean of the observed $N_p$ in Figure A1b is $(5.5\pm6.4)\times10^{-2}$/hour, being consistent with the value for the heating within error{s} (hourly mean $N_p$ is available in S2). Eq.\eqref{eq:A17}, on the other hand, predicts {$I_0$} to be {$I_0\propto G_xx$}, while the observed {$I_0$} and its peak value of $\sim0.13$ \% can be fitted by a quadratic function of {$x$} with a best-fit parameter {$d=4.3\times10^{-2}$ AU} as shown in Figure A1a. 

\begin{figure}[H]
\begin{center}
\includegraphics[natwidth=12cm,natheight=12cm]{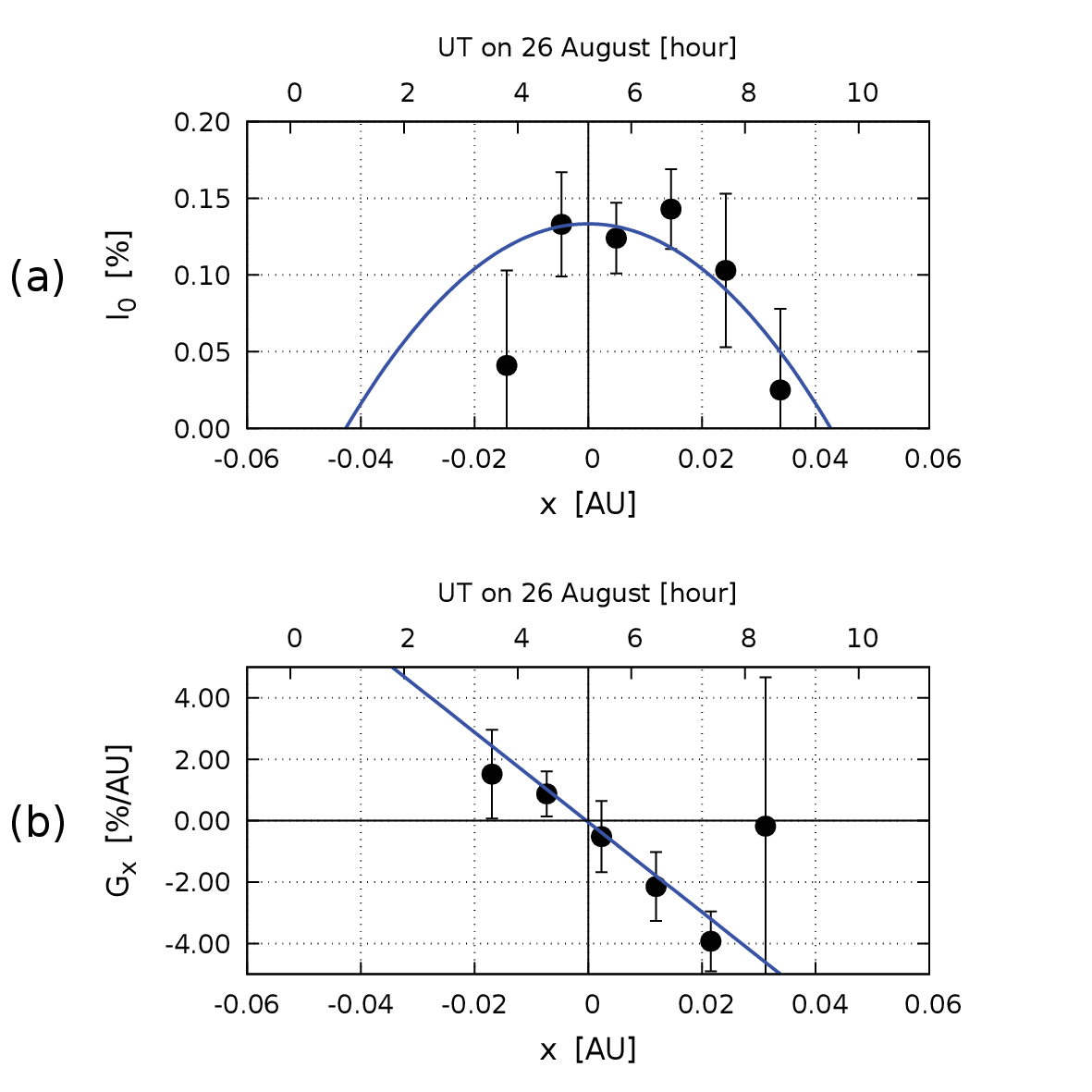}
\end{center}
\caption{Cosmic-ray density and density gradient compared with expectations from the adiabatic heating in a slab. {Each panel displays; (a) cosmic-ray density $I_0$, (b) density gradient $G_x$, as a function of $x$ measured from the center of the slab which is defined at 05:16 UT on 26 August from the zero-cross point of the best-fit curve in (b) (see the observed time on the upper horizontal axis). In this figure, we converted the time $dt$ measured from the center of the slab to the distance $x$ by $x=V_{SW}\times dt$ assuming a constant solar wind velocity of $V_{SW}=400$~km/s. The best-fit curves (blue curves) shown in panels (a) and (b) are} {$I_0 (\%) =7.3\times 10^1(\%/\rm{AU}^2)\times ((4.3\times 10^{-2}\rm{AU})^2$$-x^2)$} {and $G_x (\%/\rm{AU})=-1.47\times 10^2 (\%/\rm{AU}^2)$$\times x$ (AU)}, respectively (see text).}
\label{fig:A1}
\end{figure}

\acknowledgments
This work is supported in part by the joint research programs of the National Institute of Polar Research, in Japan, the Institute for Space-Earth Environmental Research (ISEE), Nagoya University, and the Institute for Cosmic Ray Research (ICRR), University of Tokyo. The observations are supported by Nagoya University with the Nagoya muon detector, by INPE and UFSM with the São Martinho da Serra muon detector, by the Australian Antarctic Division with the Hobart muon detector, and by project SP01/09 of the Research Administration of Kuwait University with the Kuwait City muon detector.
{Global Muon Detector Network data are available at the website (\url{http://cosray.shinshu-u.ac.jp/crest/DB/Public/main.php}) of the Cosmic Ray Experimental Science Team (CREST) of Shinshu University.}
The authors gratefully acknowledge the NOAA Air Resources Laboratory (ARL) for the provision of GDAS data, which are available at READY website (\url{http://www.ready. noaa.gov}) and used in this paper. The {\textit{Wind}} spacecraft data were obtained via the NASA homepage and the hourly {$D_{ST}$} index is provided by the WDC for Geomagnetism, Kyoto, Japan. N. J. S. thanks the Brazilian Agency - CNPq for the fellowship under grant number 300886/2016-0 and C.R.B. acknowledges grants \#2014/24711-6 and \#2017/21270-7 from S\~{a}o Paulo Research Foundation (FAPESP). EE would like to thank Brazilian funding agencies for research grants FAPESP (2018/21657-1) and CNPq (PQ-301883/2019-0).


\begin{thebibliography}{}
\bibitem[{\textit{Abunin et al.}(2020)}]{Abun20} Abunin, A. A., M. A. {Abunina}, A. V. Belov, and I. M. Chertok (2020), Peculiar Solar Sources and Geospace Disturbances on 20-26 August 2018, \textit{Solar Phys., 295}, 7, doi:10.1007/s11207-019-1574-8.
{
\bibitem[{\textit{Bieber and Evenson}(1998)}]{Bieb98} Bieber, J. W., and P. Evenson (1998), CME Geometry in Relation to Cosmic Ray Anisotropy, \textit{Geophys. Res. Lett, 25}, 2955, doi:10.1029/98GL51232.
}
\bibitem[{\textit{Bieber et al.}(2004)}]{Bieb04} Bieber, J. W., W. H. Matthaeus, and A. Shalchi (2004), Nonlinear guiding center theory of perpendicular diffusion: General properties and comparison with observation, \textit{Geophys. Res. Lett, 31}, L10805, doi:10.1029/2004GL020007.
\bibitem[{\textit{Cane}(2000)}]{Cane00} Cane, H. (2000), Coronal mass ejections and Forbush decreases, \textit{Space Sci. Rev., 93}, 55, doi:10.1023/A:1026532125747.
\bibitem[{\textit{Chen et al.}(2019)}]{Chen19} Chen, C., Y. D. Liu, R. Wang, X. Zhao, H. Hu, and B. Zhu (2019), Characteristics of a Gradual Filament Eruption and Subsequent CME Propagation in Relation to a Strong Geomagnetic Storm, \textit{Astrophys. J., 884}, 90, doi:10.3847/1538-4357/ab3f36.
\bibitem[{\textit{Dal Lago et al.}(2006)}]{Dal06} Dal Lago, A., W. D. Gonzalez, L. A. Balmaceda, L. E. A. Vieira, E. Echer, F. L. Guarnieri, J. Santos, M. R. da Silva, A. de Lucas,  A. L. C. de Gonzalez, R. Schwenn, and N. J. Schuch (2006), The 17-22 October (1999) solar-interplanetary-geomagnetic event: Very intense geomagmetic storm associated with a pressure balance between interplanetary coronal mass ejection and a high-speed stream, \textit{J. Geophys. Res., 111}, A07S14, doi:10.1029/2005JA011394.
\bibitem[{\textit{Fushishita et al.}(2010)}]{Fusi10} Fushishita, A., Y. Okazaki, T. Narumi, C. Kato, S. Yasue, T. Kuwabara, J. W. Bieber, P. Evenson, M. R. Da Silva, A. Dal Lago, N. J. Schuch, M. Tokumaru, M. L. Duldig, J. E. Humble, I. Sabbah, J. Kóta, and K. Munakata (2010), Drift effects and the average features of cosmic ray density gradient in CIRs during successive two solar minimum periods, \textit{Advances in Geosciences, 21}, 199, doi:10.1142/9789812838209\_0016.
\bibitem[{\textit{Forbush}(1937)}]{Forb37} Forbush, S. E. (1937), On the effects in cosmic-ray intensity observed during the recent magnetic storm, \textit{Phys. Rev., 51}, 1108, doi:10.1103/PhysRev.51.1108.3.
\bibitem[{\textit{Gleeson}(1969)}]{Glee69} {Gleeson, L. J. (1969), The equations describing the cosmic-ray gas in the interplanetary region, \textit{Planet. Space Sci., 17}, 31, doi:10.1016/0032-0633(69)90121-4.}
\bibitem[{\textit{Kataoka and Miyoshi}(2006)}]{Kata06} Kataoka, R., and Y. Miyoshi (2006), Flux enhancement of radiation belt electrons during geomagnetic storms driven by coronal mass ejections and corotating interaction regions, \textit{Space Weather, 4}, S09004, doi:10.1029/2005SW000211.
\bibitem[{\textit{Kataoka et al.}(2015)}]{Kata15} Kataoka, R., D. Shiota, E. Kilpua, and K. Keika (2015), Pileup accident hypothesis of magnetic storm on 2015 March 17, \textit{Geophys. Res. Lett, 42}, 5155-5161, doi:10.1002/2015GL064816.
\bibitem[{\textit{Kozai et al.}(2016)}]{Koza16} Kozai M., K. Munakata, C. Kato, T. Kuwabara, M. Rockenbach, A. Dal Lago, N. J. Schuch, C. R. Braga, R. R. S. Mendon, H. K. Al Jassar, M. M. Sharma, M. L. Duldig, J. E. Humble, P. Evenson, I. Sabbah, and M. Tokumaru (2016), Average spatial distribution of cosmic rays behind the interplanetary shock, \textit{Astrophys. J., 825}, 100, doi:10.3847/0004-637X/825/2/100.
{
\bibitem[{\textit{Krittinatham and Ruffolo}(2009)}]{Krit09} Krittinatham. W., and D. Ruffolo (2009), Drift orbits of energetic particles in an interplanetary magnetic flux rope \textit{Astrophys. J., 705}, 831, doi:10.1088/0004-637X/704/1/831.
}
\bibitem[{\textit{Kuwabara et al.}(2004)}]{Kuwa04} Kuwabara, T., K. Munakata, S. Yasue, C. Kato, S. Akahane, M. Koyama, J. W. Bieber, P. Evenson, R. Pyle, Z. Fujii, M. Tokumaru, M. Kojima, K. Marubashi, M. L. Duldig, J. E. Humble, M. Silva, N. Trivedi, W. Gonzalez, and N. J. Schuch (2004), Geometry of an interplanetary CME on October 29, 2003 deduced from cosmic rays, \textit{Geophys. Res. Lett, 31}, L19803, doi:10.1029/2004GL020803.
\bibitem[{\textit{Kuwabara et al.}(2009)}]{Kuwa09} Kuwabara, T., J. W. Bieber, P. Evenson, K. Munakata, S. Yasue, C. Kato, A. Fushishita, M. Tokumaru, M. L. Duldig, J. E. Humble, M. R. Silva, A. Dal Lago, and N. J. Schuch (2009), Determination of ICME Geometry and Orientation from Ground Based Observations of Galactic Cosmic Rays, \textit{J. Geophys. Res., 114}, A05109, doi:10.1029/2008JA013717.
\bibitem[{\textit{Liu et al.}(2014)}]{Liu14} Liu, Y., J. Luhmann,  P. Kajdič, E. K. J. Kilpua, N. Lugaz, N. V. Nitta, C. M\"{o}stl, B. Lavraud, S. D. Bale, C. J. Farrugia, and A. B. Galvin (2014), Observations of an extreme storm in interplanetary space caused by successive coronal mass ejections, \textit{Nat Commun 5}, 348, doi:10.1038/ncomms4481.
\bibitem[{\textit{Matsumoto et al.}(2019)}]{Matu19} Matsumoto, T., D. Shiota, R. Kataoka, H. Miyahara, and S. Miyake (2019), A dynamical model of the heliosphere with the Adaptive Mesh Refinement, \textit{Phys.: Conf. Ser., 1225}, 1225, 012008, doi:10.1088/1742-6596/1225/1/012008.
\bibitem[{\textit{Mendonça et al.}(2016)}]{Rafa16} Mendonça, R. R. S., C. R. Braga, E. Echer, A. Dal Lago, K. Munakata, T. Kuwabara, M. Kozai, C. Kato, M. Rockenbach, N. J. Schuch, H. K. Al Jassar, M. M. Sharma, M. Tokumaru, M. L. Duldig, J. E. Humble, P. Evenson, and I. Sabbah (2016), Temperature effect in secondary cosmic rays (muons) observed at ground: analysis of the global muon detector network data, \textit{Astrophys. J., 830}, 88, doi:10.3847/0004-637X/830/2/88.
\bibitem[{\textit{Miyake et al.}(2017)}]{Miya17} Miyake, S., R. Kataoka, and T. Sato (2017), Cosmic ray modulation and radiation dose of aircrews during the solar cycle 24/25, \textit{ Space Weather, 15(4)},589-605, doi:0.1002/2016SW001588
\bibitem[{\textit{Mohanty et al.}(2016)}]{Moha16} Mohanty, P. K., K. P. Arunbabu, T. Aziz, S. R. Dugad, S. K. Gupta, B. Hariharan, P. Jagadeesan, A. Jain, S. D. Morris, B. S. Rao, Y. Hayashi, S. Kawakami, A. Oshima, S. Shibata, S. Raha, P. Subramanian, and H. Kojima (2016), Transient Weakening of Earth's Magnetic Shield Probed by a Cosmic Ray Burst, \textit{Phys. Rev. Lett., 117}, 171101, doi:10.1103/PhysRevLett.117.171101.
\bibitem[{\textit{Munakata et al.}(2006)}]{Muna06} Munakata, K., S. Yasue, C. Kato, J. Kota, M. Tokumaru, M. Kojima, A. A. Darwish, T. Kuwabara, and J. W. Bieber (2006), On the cross-field diffusion of galactic cosmic rays into the magnetic flux rope of a CME,  \textit{Advances in Geosciences, 21}, 115, doi:10.1142/9789812707185.
\bibitem[{\textit{Munakata et al.}(2018)}]{Muna18} Munakata, K., M. Kozai, P. Evenson, T. Kuwabara, C. Kato, M. Tokumaru, M. Rockenbach, A. Dal Lago, R. R. S. Mendonca, C. R. Braga, N. J. Schuch, H. K. Al Jassar, M. M. Sharma, M. L. Duldig, J. E. Humble, I. Sabbah, and J. Kota (2018), Cosmic Ray Short Burst Observed with the Global Muon Detector Network (GMDN) on June 22, 2015, \textit{Astrophys. J., 862}, 170, doi:10.3847/1538-4357/aacdfe.
\bibitem[{\textit{Murakami et al.}(1979)}]{Mura79} Murakami, K., K. Nagashima, S. Sagisaka, Y. Mishima, and A. Inoue (1979), Response Functions for Cosmic-Ray Muons at Various Depths Underground, \textit{IL NUOVO CIM., 2C}, 635, doi:10.1007/BF02557762.
\bibitem[{\textit{Okazaki et al.}(2008)}]{Oka08} Okazaki, Y., A. Fushishita, T. Narumi, C. Kato, S. Yasue, T. Kuwabara, J. W. Bieber, P. Evenson, M. R. Da Silva, A. Dal Lago, N. J. Schuch, Z. Fujii, M. L. Duldig, J. E. Humble, I. Sabbah, J. Kóta, and K. Munakata (2008), Drift effects and the cosmic ray density gradient in a solar rotation period: First observation with the Global Muon Detector Network (GMDN), \textit{Astrophys. J., 681}, 693, doi:10.1086/588277.
\bibitem[{\textit{Parker}(1965)}]{Park65} Parker, E. N. (1965), The passage of energetic charged particles through interplanetary space, \textit{Planet. Space Sci., 13}, 9, doi:10.1016/0032-0633(65)90131-5.
\bibitem[{\textit{Richardson}(2004)}]{Rich04} Richardson, I. G. (2004), Energetic particles and corotating interaction regions in the solar wind, \textit{Space Sci Rev., 111}, 267, doi:10.1023/B:SPAC.0000032689.52830.3e.
\bibitem[{\textit{Rockenbach et al.}(2014)}]{Marl14}	Rockenbach, M., A. Dal Lago, N. J. Schuch, K. Munakata, T. Kuwabara, A. G. Oliveira, E. Echer, C. R. Braga, R. R. S. Mendonça, C. Kato, M. Kozai, M. Tokumaru, J. W. Bieber, P. Evenson, M. L. Duldig, J. E. Humble, H. K. Al Jassar, M. M. Sharma, and I. Sabbah (2014), Global muon detector network used for space weather applications, \textit{Space Sci Rev., 182}, 1, doi:10.1007/s11214-014-0048-4.
\bibitem[{\textit{Shiota and Kataoka}(2016)}]{Shio16} Shiota, D., and R. Kataoka (2016), Magnetohydrodynamic simulation of interplanetary propagation of multiple coronal mass ejections with internal magnetic flux rope (SUSANOO-CME), \textit{Space Weather, 14}, 56-75, doi:10.1002/2015SW001308.
\bibitem[{\textit{Tokumaru et al.}(2007)}]{Toku07} Tokumaru, M., M. Kojima, K. Fujiki, M. Yamashita, and B. V. Jackson (2007), The source and propagation of the interplanetary disturbance associated with the full-halo coronal mass ejection on 28 October 2003, \textit{J. Geophys. Res., 112}, A05106, doi:10.1029/2006JA012043.
{\bibitem[{\textit{Tortermpun et al.}(2018)}]{Tort18} Tortermpun, U., D. Ruffolo, and J. {W.} Bieber (2018), Galactic cosmic-ray anisotropy during the Forbush decrease starting 2013 April 13, \textit{Astrophys. J. Lett., 852}, L26, doi:10.3847/2041-8213/aaa407.}
\bibitem[{\textit{{Wibberenz et al.}}(1998)}]{Wibb98} {Wibberenz}, G., J. A. Le Roux, M. S. Potgieter, and J. W. {Bieber} (1998), Transient effects and disturbed conditions, \textit{Space Sci. Rev., 83}, 309, doi:10.1023/A:1005083109827.
\end{thebibliography}
\end{document}